\documentclass[paper]{JHEP}
\usepackage{epsfig} 
\usepackage{amssymb}

\newcommand     \MSB            {\ifmmode {\overline{\rm MS}} \else
                                 $\overline{\rm MS}$  \fi}
\newcommand\LambdaQCD{\Lambda_{\rm \scriptscriptstyle QCD}}
\newcommand\LambdaQCDfive{\Lambda_{\rm \scriptscriptstyle QCD}^{(5)}}
\newcommand\muF{\mu_{\rm \scriptscriptstyle F}}
\newcommand\muR{\mu_{\rm \scriptscriptstyle R}}

\newcommand\as{\alpha_{\rm s}}
\newcommand\ep{\epsilon}
\newcommand\pt{p_{\rm\scriptscriptstyle T}}
\newcommand\aem{\alpha_{\rm em}}

\preprint{
        Bicocca-FT-02-1\hfill\\
        LAPTH-897/02\hfill\\
        GEF-th-1/2002\hfill\\
        hep-ph/0201281}
\title{Phenomenological study of charm photoproduction at HERA}
\author{Stefano Frixione%
\thanks{On leave of absence from INFN, Sez. di Genova, Italy}\\
Laboratoire d'Annecy-le-Vieux de Physique de Particules\\
Chemin de Bellevue, BP 110,
74941 Annecy-le-Vieux CEDEX, France}
\author{Paolo Nason\\
INFN, Sezione di Milano\\
Via Celoria 16, 20133 Milan, Italy}
\abstract{
  We present predictions for single inclusive distributions of charmed
  mesons, relevant to the HERA experiments. Our results are based upon
  a computation that correctly incorporates mass effects up to the
  next-to-leading order level, and the resummation of transverse
  momentum logarithms up to next-to-leading-logarithmic level.  We
  apply the same acceptance cuts as the H1 and Zeus experiments, and
  compare our results to their data. We perform a study of the
  sensitivity of our predictions on the charm mass, $\LambdaQCD$,
  factorization scale, renormalization scale, and fragmentation
  parameters.}
\keywords{QCD, NLO Computations, Resummations, Hadron Colliders, 
Heavy Quark Physics}

\begin{document}

\section{Introduction}\label{sec:intro}
In the present work we discuss phenomenological applications
of the formalism presented in ref.~\cite{Cacciari:2001td}, relevant to the
production of heavy quarks in photon-hadron collisions.
All technical details and theoretical motivations for the formalism
are discussed there.
In the formalism of ref.~\cite{Cacciari:2001td} one obtains
a cross section which is accurate to ${\cal O}(\aem\as^2)$
(that is, to the next-to-leading order level)
and which, in the large $\pt$ region, includes consistently
all enhanced logarithmic terms of the form $\aem\as (\as\log\pt/m)^i$
and $\aem\as^2 (\as\log\pt/m)^i$. Terms of order
${\cal O}(\aem\as^3)$ with no logarithmic enhancement and terms
of order $\aem\as^3 (\as\log\pt/m)^i$ are not included. 
Thus this approach improves
both the fixed order (improperly called massive) and the resummed
(improperly called massless) calculations.

Our starting point is the ``matched'' formula~\cite{Cacciari:2001td}
\begin{equation}
  \label{eq:merge}
  \mbox{FONLL}=\mbox{FO}\;+\left( \mbox{RS}\; -\; \mbox{FOM0}\right)\;
\times G(m,\pt)\;,
\end{equation}
where FONLL stands for fixed-order plus next-to-leading logarithms, and
\begin{itemize}
\item FO is the fixed order, ${\cal O}(\aem\as^2)$ result;
\item RS is the resummed result, which includes all terms of the
form  $\aem\as (\as\log\pt/m)^i$
and $\aem\as^2 (\as\log\pt/m)^i$, and neglects all terms suppressed
by powers of the heavy quark mass $m$;
\item FOM0 is the massless limit of FO, in the sense that all terms
suppressed by powers of $m$ are dropped, while logarithms of the
mass are retained; thus FOM0 is the truncation of RS to
order $\aem\as^2$;
\item  $G(m,\pt)$ is an arbitrary dumping function, that must be
regular in $\pt$, and must approach 1 up to terms suppressed by
powers of $m/\pt$ at large $\pt$; our standard choice is
\begin{equation}
G(m,\pt)=\frac{\pt^2}{\pt^2 + c^2 m^2}\;,
\label{gfundef}
\end{equation}
which for $c\ne 0$ dumps the $\mbox{RS}\; -\; \mbox{FOM0}$ term
in the small-$\pt$ region.
\end{itemize}
We remind the reader that in the description given above the
hadronic photon contribution is also included; the counting
of the orders in $\aem$ and $\as$ remains the same, provided
one counts the parton density functions of the photon
as carrying a $\aem/\as$ factor. A more detailed discussion
can be found in ref.~\cite{Cacciari:2001td}.

This paper is organized as follows. In sect.~\ref{sec:inputs}
we specify our choices for the input parameters and the parton
density functions (PDF's). Furthermore, the specific cuts that the
Zeus and H1 collaborations use in their analyses are reported
in this section.  All the subsequent results will be based upon
these choices.

In order to present meaningful phenomenological results, we need to
consider a non-perturbative fragmentation function (NPFF) to account
for the hadronization of the charm quark into a charmed hadron.  The
NPFF is the source of several uncertainties affecting the theoretical
predictions. First of all, various forms of NPFF have been proposed in
the literature (see ref.~\cite{Biebel:2001ka} and references therein
for an updated review). Furthermore, while the application of NPFF is
unambiguous in the large $\pt$ region, at moderate and small $\pt$
there are ambiguities, depending upon the definition of the $z$
variable (i.e., whether it refers to energy, momentum or ``plus''
component), and upon the choice of the frame in which it is defined.
The impact of these ambiguities will be assessed in
sect.~\ref{sec:npff}.

Phenomenological results have also uncertainties depending upon the
accuracy of the perturbative calculations (which we estimate by
varying the renormalisation and factorisation scales), 
the allowed range in the value of the heavy quark mass and
$\LambdaQCD$, and the choice of the PDF's. These uncertainties will be
discussed in sect.~\ref{sec:range}.

The H1~\cite{Adloff:1998vb} and the Zeus~\cite{Breitweg:1998yt}
collaborations have studied photoproduction of $D^*$ mesons. The
two experiments have different acceptances and apply different cuts
to their data samples. In sect.~\ref{sec:data} we compute the
distributions presented in refs.~\cite{Adloff:1998vb,Breitweg:1998yt},
implementing the same cuts that are used there.

In sect.~\ref{sec:conc} we give our conclusions.

\section{Input parameters}\label{sec:inputs}
All the results presented in this paper are relevant to photoproduction
experiments where the photon arises from collinear radiation off an
incoming electron. The radiation spectrum depends upon the cuts that
are applied to define the photoproduction regime, which are
typically $Q^2$ cuts. Further kinematic cuts are applied either to the
the variable $y$ (the ratio of the photon energy over the incoming
electron energy) or, which is equivalent, to the variable $W$ (the
photon-proton center-of-mass energy).
The cuts used in refs.~\cite{Adloff:1998vb,Breitweg:1998yt}
are summarised in table~\ref{tab:q2ycuts}.
\begin{table}[htb]
\begin{center}
\begin{tabular}{|c|c|c|c|}\hline
      & $Q_{\rm max}^2$ (GeV$^2$) & $y_{\rm min}$ &  $y_{\rm max}$  \\\hline
Zeus      & $1$             & $0.187$       &  $0.869$        \\\hline
H1 ETAG44 & $0.009$         & $0.02$        &  $0.32$         \\\hline
H1 ETAG33 & $0.01$          & $0.29$        &  $0.62$         \\\hline
\end{tabular}
\end{center}
\caption{\label{tab:q2ycuts}
        $Q^2$ and $y$ cuts used by H1 and Zeus in 
        refs.~\cite{Adloff:1998vb,Breitweg:1998yt}
        to define a photoproduction event.}
\end{table}
The H1 collaboration uses two different electron taggers, denoted as
ETAG33 and ETAG44, and has thus two samples of photoproduction data.
The electron and proton energies are fixed to $27.5$ and $820$~GeV
respectively. The $y$ range quoted for Zeus corresponds to
$130<W<280$~GeV.

We compute the photon spectrum in the Weizs\"acker-Williams approximation,
in the improved form of ref.~\cite{Frixione:1993yw}, which includes
non-logarithmic terms enhanced at small $y$.

The central values and ranges of the physical parameters that we adopt
in the present paper are summarised in table~\ref{tab:physpar}.  The
range in $\LambdaQCDfive$ reported in the table corresponds to the
range $\as(M_{\rm Z})=0.118 \pm 0.002$.  We use the two-loop
expression of $\as$ throughout this work.
\begin{table}[htb]
\begin{center}
\begin{tabular}{|c|c|c|c|} \hline
            & low & central & high  \\\hline
Proton PDF  & \multicolumn{3}{|c|}{CTEQ5M~\cite{Lai:1999wy}}  \\\hline
Photon PDF  & \multicolumn{3}{|c|}{AFG~\cite{Aurenche:1994in}}     \\\hline
Charm quark mass $m$ & $1.2$~GeV & $1.5$~GeV & $1.8 $~GeV   \\\hline
$\LambdaQCDfive$ & $0.203$        & $0.226$       &  $0.254$        \\\hline
\end{tabular}
\end{center}
\caption{\label{tab:physpar}
Central values and ranges of the physical parameters adopted
in this paper.}
\end{table}
The factorization and renormalization scales ($\muF$ and $\muR$)
are set equal to $\mu_0/2$, $\mu_0$ and $2\mu_0$, where
$\mu_0=\sqrt{\pt^2+m^2}$ is our default choice.

In table~\ref{tab:visible} we report the kinematic ranges
of the cross section measurements used in the Zeus and H1 analyses.
Notice that the $D^*$ kinematics are described by $\pt,\eta$ (pseudorapidity)
in the Zeus analysis, and by $\pt,y$ (rapidity) in the H1 analysis.
Unfortunately, the same notation ($y$) is traditionally used to indicate
the Weizs\"acker-Williams variable, and the meson rapidity. However, no
confusion should possibly arise. In what follows, $y$ will always
indicate the meson rapidity.
\begin{table}[htb]
\begin{center}
\begin{tabular}{|c|c|c|}\hline
Zeus      & $\pt(D^*)>2$~GeV   & $|\eta(D^*)|<1.5$ \\\hline
H1 ETAG44 & $\pt(D^*)>2$~GeV   & $|y(D^*)|<1.5$    \\\hline
H1 ETAG33 & $\pt(D^*)>2.5$~GeV & $|y(D^*)|<1.5$    \\\hline
\end{tabular}
\end{center}
\caption{\label{tab:visible}
Visible ranges for the H1 and Zeus analyses of 
refs.~\cite{Adloff:1998vb,Breitweg:1998yt}.}
\end{table}

\section{Non-perturbative fragmentation function}\label{sec:npff}
We begin by discussing the ambiguities related to the implementation
of the non-perturbative fragmentation in our calculation.
A NPFF has to be convoluted with our full FONLL result.
The convolution is unambiguously defined only at large transverse
momenta (i.e. $\pt\gg m$). One writes
\begin{equation}\label{NPFFxsec}
\frac{d^3\sigma_H(k)}{d^3 k} = \int d^3 \hat{k} \;dz\,D_{\rm NP}(z)
\frac{d^3\sigma_Q(\hat{k})}{d^3 \hat{k}}
\delta^3\left(\vec{k}-z\vec{\hat{k}}\right)\;,
\end{equation}
where $\sigma_H$ is the cross section for the production of 
the heavy-flavoured hadron\footnote{In what follows, we assume 
the hadron $H$ to be a $D^*$ meson, since the H1 and Zeus data
that we use in this paper are relevant to $D^*$ mesons.}
$H$ with momentum $k$, and $\sigma_Q$ is the cross section for the production
of the heavy quark $Q$ with momentum $\hat{k}$. $D_{\rm NP}(z)$ is the NPFF. 
In eq.~(\ref{NPFFxsec}) we assume that fragmentation scales the
3-momentum of the heavy quark.  Different prescriptions are however
possible: for example, one could assume that either $\hat{k}^0$ or
$\hat{k}^0+|\hat{k}|$ are scaled instead of $\vec{k}$.  In all cases,
it is assumed that $\vec{k}$ remains parallel to $\vec{\hat{k}}$, and
the mass-shell condition is imposed to fix the remaining components of
the meson momentum.  Only in the large-$\pt$ limit all these
prescriptions coincide.  Also, none of these definitions is boost
invariant, since the condition of parallelism is frame dependent.
Again, in the large-$\pt$ limit boost invariance is recovered.

For the meson momentum $k$ we expect the condition
$k^2=m_H^2$ to be satisfied. However, for the purpose of computing
distributions, we can also use $k^2=m^2$, and consider the difference
to be just another ambiguity in our result. The reader may find
unreasonable to adopt this second choice, since physically the
on-shell condition of the meson $H$ must apply.
However, in the context of a perturbative
QCD computation, setting $k^2=m_H^2$ may appear unreasonable from other
points of view. For example, if we produce a quark with maximal $\pt$, and
fragment it with $z=1$, setting $k^2=m_H^2$ would yield a meson with the same
$\pt$ but larger energy, which violates energy conservation.
Similar examples can be given
at small $\pt$ and large rapidities. If we are far from the phase space
boundaries, which is always our case, both choices are sensible,
and their difference can be considered as an inherent uncertainty of the
procedure.

In the following, we will always use the Peterson fragmentation function
\cite{Peterson:1983ak}
\begin{equation}\label{peterson}
D_{\rm NP}(z)\propto \frac{1}{z}
      \left(1-\frac{1}{z}-\frac{\epsilon}{1-z}\right)^{-2}\,.
\end{equation}
For the sake of the studies performed in the present section we will
take it normalized to 1, i.e. we do not include the $D^*$ production
fraction.  Although other choices of fragmentation function are
possible, we only use eq.~(\ref{peterson}) because it is the most
commonly adopted, and because recent phenomenological fits of the
$\epsilon$ parameter, adequate to the present work, do exist
\cite{Cacciari:1997du,Nason:1999zj}.
In refs.~\cite{Biebel:2001ka,Nason:1999ta} 
the reader will find studies of alternative forms.

Our default choices for the non-perturbative fragmentation are as follows:
\begin{itemize}
\item the $z$ variable is defined by the equation $\vec{k}=z\vec{\hat{k}}$;
\item the fragmentation frame (i.e., the frame where the equation 
$\vec{k}=z\vec{\hat{k}}$ holds) is the HERA laboratory frame;
\item the mass shell condition is $k^2=\hat{k}^2=m^2$;
\item the value of the Peterson parameter is $\epsilon=0.02$.
\end{itemize}

We now study the effect of the ambiguities in the definition of the
non-perturbative fragmentation on the pseudorapidity
and on the transverse momentum distributions for the Zeus kinematic
cuts. Our conclusions would not change had we used H1 cuts.
We consider first the effect of the change of the fragmentation frame.
We call scheme (A) our default choice of frame.
We call scheme (B) the alternative choice of a longitudinally-%
boosted fragmentation frame, boosted with rapidity equal to the quark rapidity.
Thus, in this frame the quark rapidity is zero.
We notice that in scheme (A) the quark and the
meson have the same pseudorapidity, while in scheme (B) they have the
same rapidity.

\FIGURE[htb]{
    \epsfig{figure=ZeusNPFFeta.eps,width=0.49\textwidth}
    \epsfig{figure=ZeusNPFFpt.eps,width=0.49\textwidth}
    \caption{\label{fig:ZeusNPFF} \protect\small
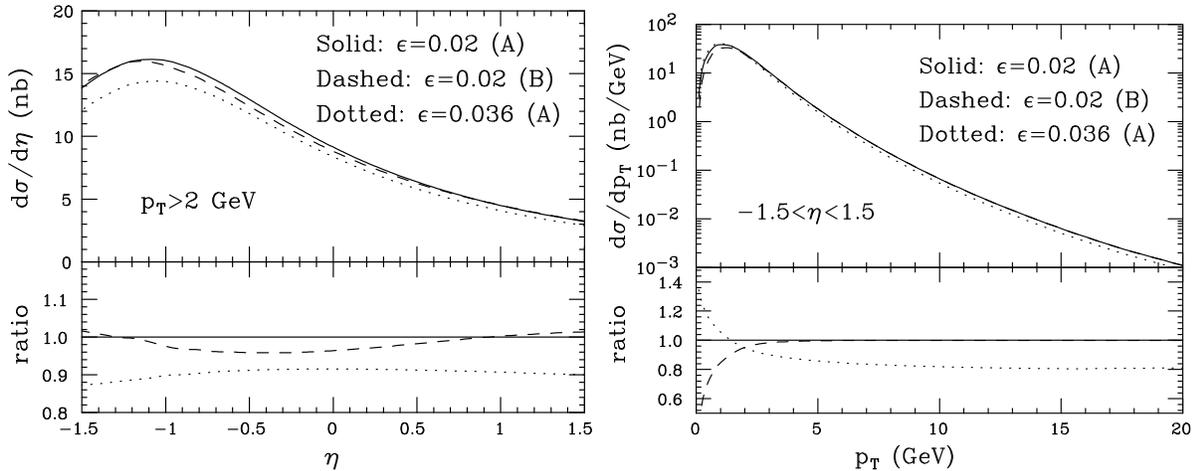
       Fragmentation scheme and $\ep$ dependence of
       the pseudorapidity and transverse momentum distributions with
       Zeus cuts. The lower plots show the ratio of the variation
       over the default choice.}}
In figure~\ref{fig:ZeusNPFF} we show the
difference in the pseudorapidity and transverse momentum distributions
when using schemes (A) and (B). Furthermore, the effect of varying
the value of the Peterson parameter $\ep$ (in scheme (A)) is also shown.
As we can see, the dependence on the fragmentation scheme is visible
only for $\pt < 3\;$GeV. The rapidity distribution is mostly affected because
of the relatively low $\pt$ cut. The effect is generally modest,
below 5\%\ in the range of interest.
When increasing $\epsilon$, the $\pt$ spectrum becomes softer.
The spectrum with $\epsilon=0.036$ is below the one with
$\epsilon=0.02$ for $\pt>1.5$~GeV; in the tail,
the difference is of the order of 30\%. As in the case of
the fragmentation frame dependence, the behaviour of the $\pt$ 
spectrum affects the $\eta$ distribution through the $\pt$ cut.
When changing $\epsilon$ from 0.02 to 0.036, the $\eta$ spectrum 
is almost unchanged in shape, and the normalisation decreases of about 10\%.

\FIGURE[htb]{
    \epsfig{figure=ZeusNPFFetaMh.eps,width=0.49\textwidth}
    \epsfig{figure=ZeusNPFFptMh.eps,width=0.49\textwidth}
    \caption{\label{fig:ZeusNPFFMh} \protect\small
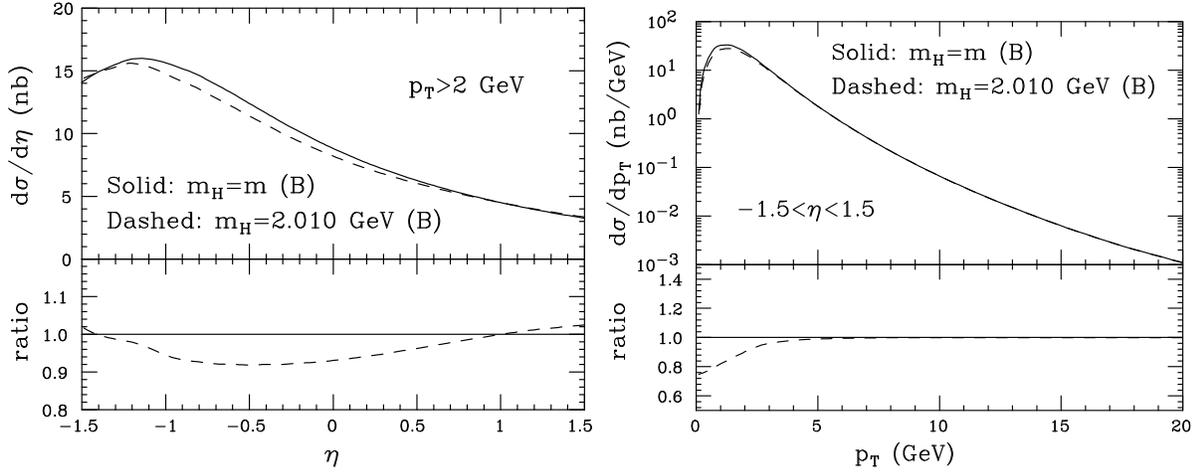
       Effect of the on-mass-shell prescription in
       the pseudorapidity and transverse momentum distributions with
       Zeus cuts. The lower plots show the ratio of the variation
       over the default choice.}}
In figure~\ref{fig:ZeusNPFFMh} we show the
difference in the pseudorapidity and transverse momentum distributions
when using the mass of the quark (our default choice) for the on-mass-shell
condition, as opposed to the mass of the $D^*$ meson ($m_{D^*}=2.010\;$GeV).
The comparison is performed in scheme (B), since in scheme (A) there is
no difference in $\pt$ and $\eta$ for the two mass choices.
The effect is only visible below $\pt=5\;$GeV. In the case of HERA
experiments, only $\pt>2\;$GeV is relevant; 
in this region, the variation is below 10\%. Furthermore, larger
relative variations are observed only for very small $\pt$,
where the differential cross section is actually smaller in absolute
value.

\section{Dependence upon input parameters}\label{sec:range}
We now turn to the study of the sensitivity of our result to the
chosen range of parameters. We begin by showing in fig.~\ref{fig:Zeus-lambda}
the sensitivity to the choice of $\LambdaQCD$.
\FIGURE[htb]{
    \epsfig{figure=Zeus-pt-lambda.eps,width=0.49\textwidth}
    \epsfig{figure=Zeus-eta-lambda.eps,width=0.49\textwidth}
    \caption{\label{fig:Zeus-lambda} \protect\small
       Sensitivity of the transverse momentum and pseudorapidity distributions
       to the choice of $\LambdaQCD$.}}
As we can see the sensitivity is quite small, below 10\%. This
is a consequence of the fact that the allowed range for $\LambdaQCD$
has become fairly narrow at present. 

We do not include in our study the correlation of $\LambdaQCD$ with
the parton densities, since PDF sets with flavour matching conditions
consistent with our scheme that allow for $\LambdaQCD$ variation are
not available at present. For the same reason, we do not include a
study of the uncertainty due to the choice of the PDF set. In
general, the inclusion of such an uncertainty is a highly non trivial
problem, and it is far from having a satisfactory solution at present.
However, it must be said that current PDF sets have rather similar
gluon densities in the $x$ region probed by charm production at HERA; 
therefore, we do not expect the PDF dependence to be a major source
of uncertainty in our case.

\FIGURE[htb]{
    \epsfig{figure=Zeus-pt-mass.eps,width=0.49\textwidth}
    \epsfig{figure=Zeus-eta-mass.eps,width=0.49\textwidth}
    \caption{\label{fig:Zeus-mass} \protect\small
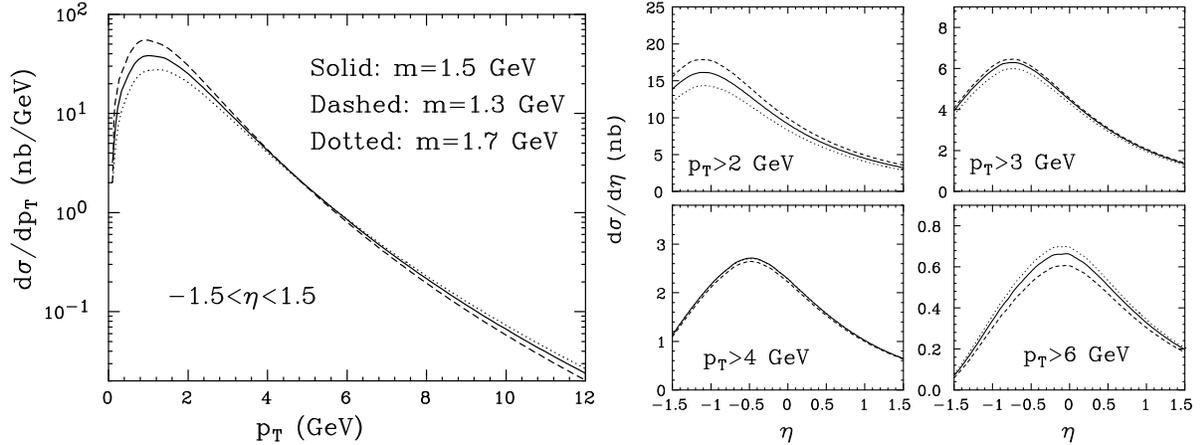
       Sensitivity of the transverse momentum and pseudorapidity distributions
       to the choice of the charm mass.}}
The mass sensitivity is shown in fig.~\ref{fig:Zeus-mass}.
The cross section is most sensitive to the mass value for $\pt$ of the order 
of 1~GeV. In the visible range ($\pt>2$~GeV) the sensitivity is modest,
of the order of 10{\%}. 

It has to be pointed out that our study of mass sensitivity is
somewhat incomplete. In fact, we do not have the possibility of
varying the charm mass in the evolution of the parton densities.
Studies performed in the hadroproduction case have indicated that
these effects are small.

\FIGURE[htb]{
    \epsfig{figure=Zeus-pt-rensc.eps,width=0.49\textwidth}
    \epsfig{figure=Zeus-eta-rensc.eps,width=0.49\textwidth}
    \caption{\label{fig:Zeus-rensc} \protect\small
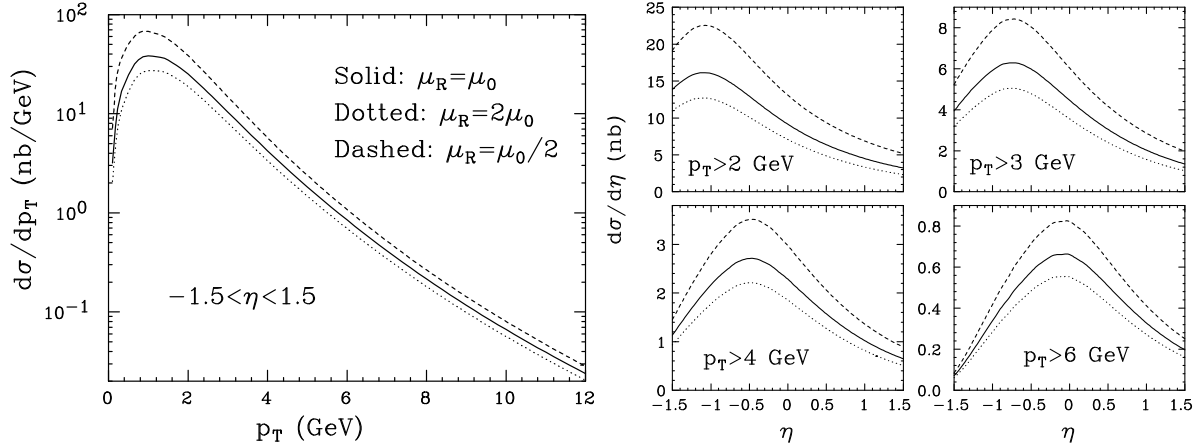
       Sensitivity of the transverse momentum and pseudorapidity distributions
       to the choice of the renormalization scale.}}
\FIGURE[htb]{
    \epsfig{figure=Zeus-pt-factsc.eps,width=0.49\textwidth}
    \epsfig{figure=Zeus-eta-factsc.eps,width=0.49\textwidth}
    \caption{\label{fig:Zeus-factsc} \protect\small
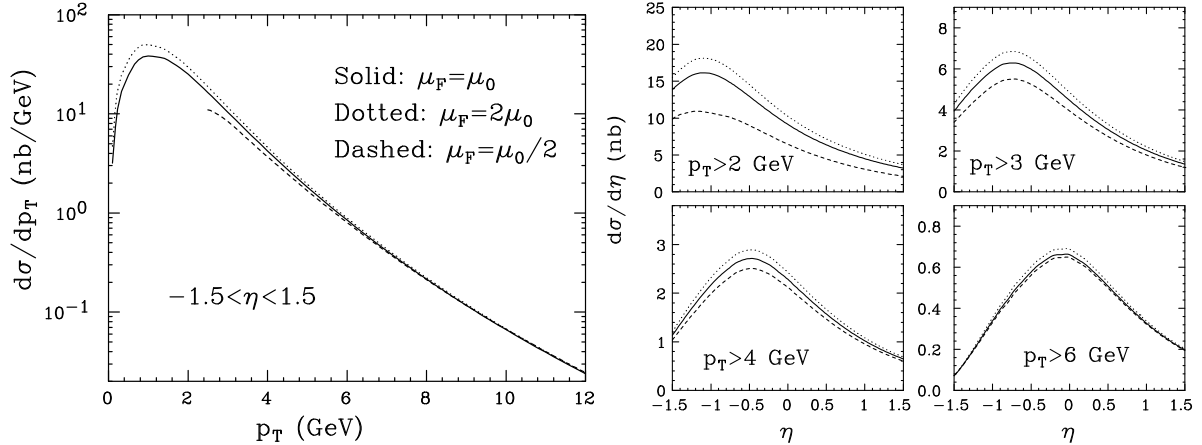
       Sensitivity of the transverse momentum and pseudorapidity distributions
       to the choice of the factorization scale.}}
The renormalization and factorization scale dependences are displayed in
figures~\ref{fig:Zeus-rensc} and \ref{fig:Zeus-factsc} respectively.
The factorization scale choice has little impact on our results,
except for small transverse momenta. In particular, when
$\muF=\mu_0/2$, at small $\pt$ the cross section becomes unphysical,
since it depends upon the parton densities at low scales,
a region where current parametrizations are unreliable.
We simply truncate the theoretical predictions
at small transverse momenta in this case.
The renormalization scale dependence is instead sizable in the
whole production range. Since no data are available for $\pt<2$~GeV,
the inspection of figures~\ref{fig:Zeus-lambda}-\ref{fig:Zeus-factsc}
allows us to conclude that the renormalization scale dependence is the 
largest source of uncertainty when comparing the theoretical results 
to the data.

\section{Comparison with data}\label{sec:data}
In this section, we compare the theoretical predictions, obtained with
our FONLL formalism, with the data relevant to $D^*$ production,
presented by the H1 and Zeus collaborations in
refs.~\cite{Adloff:1998vb,Breitweg:1998yt}. For the sake of this
comparison, we now need to multiply the NPFF of eq.~(\ref{peterson})
by a factor of 23.5~\%, to take into account the $c\to D^*$
fraction~\cite{Gladilin:1999pj}. Furthermore, we multiply by a factor of 2
in order to get the ${D^*}^+ + {D^*}^-$ cross section,
which is what the experiments quote in 
refs.~\cite{Adloff:1998vb,Breitweg:1998yt}. H1 data are presented
in ref.~\cite{Adloff:1998vb} as $\gamma p$ cross sections, obtained 
by dividing the measured photoproduction $ep$ cross sections by the 
Weizs\"acker-Williams flux; here, we quote the data as $ep$ cross sections.

In QCD, it is very often the case that the theoretical uncertainties
are larger than the experimental errors. Thus, an estimate of these
uncertainties is mandatory for a meaningful comparison with the data.
Usually, one takes the maximum and minimum values of the theoretical
predictions, in the allowed range of the input parameters, as an
estimate of the upper and lower theoretical errors. In our case, since
the computation is numerically intensive, this procedure is too time
consuming. We have thus adopted the following procedure.  We construct
an error band for our predictions by adding linearly all the
variations that we have discussed in the previous section. In other
words, for each parameter $x$ (mass, $\LambdaQCD$, $\muR$, $\muF$)
that has central value $x_0$ and upper and lower values $x_h$ and
$x_l$ respectively, we define an upper and lower error
\begin{eqnarray}
\delta^{\rm up}_x &=&
   {\rm max}[\sigma(x_0),\sigma(x_h),\sigma(x_l)]-\sigma(x_0)\,,
\label{deltaup} \\
\delta^{\rm down}_x &=&\sigma(x_0)-
 {\rm min}[\sigma(x_0),\sigma(x_h),\sigma(x_l)]\,,
\label{deltadown}
\end{eqnarray}
with all the other parameters in the cross section kept fixed
to their central values. We then define our upper (lower) end of the band as 
the central cross section plus (minus) the sum of all upper (lower) errors.
By interpreting eqs.~(\ref{deltaup}) and~(\ref{deltadown}) as true
errors affecting the theoretical predictions, we could have also
adopted the strategy of adding them quadratically. We refrain from doing
this, since it is actually very doubtful that they obey a Gaussian law.
However, it must be stressed that, in practice, the quadratic sum
differs only by a tiny amount from the linear one, since they are
both dominated by the effect due to the renormalization scale,
as discussed in the previous section.

\FIGURE[htb]{
    \epsfig{figure=Zeus-pt-band.eps,width=0.49\textwidth}
    \epsfig{figure=Zeus-eta-band.eps,width=0.49\textwidth}
    \caption{\label{fig:Zeus-Band} \protect\small
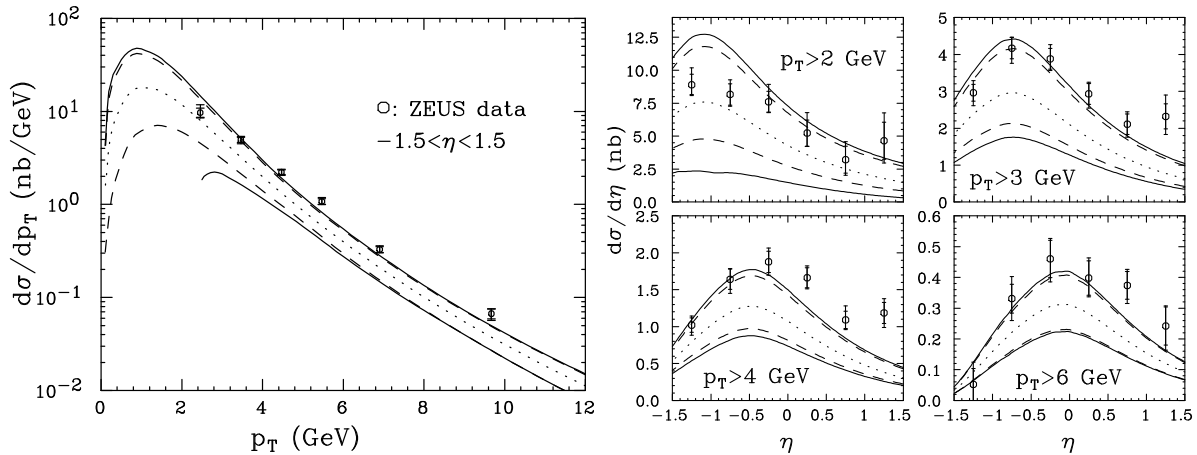
       Comparison of Zeus data with the theoretical predictions.
       The dotted curve is the central value, the band included
       between the solid curves is obtained by adding linearly the
       uncertainties, the dashed band does not include the
       factorization scale uncertainty.
}}
The lower error due to the factorization scale dependence is undefined for 
small momenta. Thus, the lower end of the band is undefined
for low momenta if the factorization scale dependence is included
in the error. We therefore present one band that includes the factorization 
scale error, and one band that does not include it. Since the factorization 
scale dependence is in general rather small, the difference between the
two bands is non-negligible only in the small-$\pt$ region, basically outside
the region visible to experiments.

We begin with Zeus data from ref.~\cite{Breitweg:1998yt}, for which we
present a comparison with our FONLL predictions in
fig.~\ref{fig:Zeus-Band}.  As often stated in the literature, the data
lie above the theoretical predictions obtained with the default choice
of parameters. On the other hand, we see that the data are marginally
consistent with our upper band. However, the data seem to indicate a
$\pt$ spectrum harder than that suggested by QCD.  As far as the
pseudorapidity distribution is concerned, QCD appears to do a decent
job, except in the positive-$\eta$ region. The rightmost data points,
however, are by far the ones affected by the largest errors.

\FIGURE[htb]{
    \epsfig{figure=h33-pt-band.eps,width=0.49\textwidth}
    \epsfig{figure=h33-y-band.eps,width=0.49\textwidth}
    \caption{\label{fig:H33-Band} \protect\small
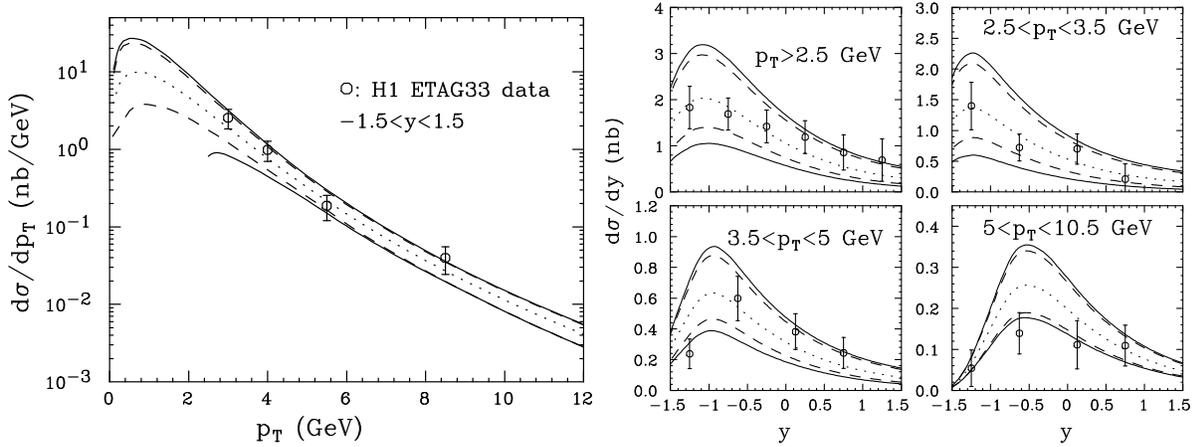
       As in fig.~\ref{fig:Zeus-Band}, for the H1 ETAG33 data.
}}
\FIGURE[htb]{
    \epsfig{figure=h44-pt-band.eps,width=0.505\textwidth}
    \epsfig{figure=h44-y-band.eps,width=0.475\textwidth}
    \caption{\label{fig:H44-Band} \protect\small
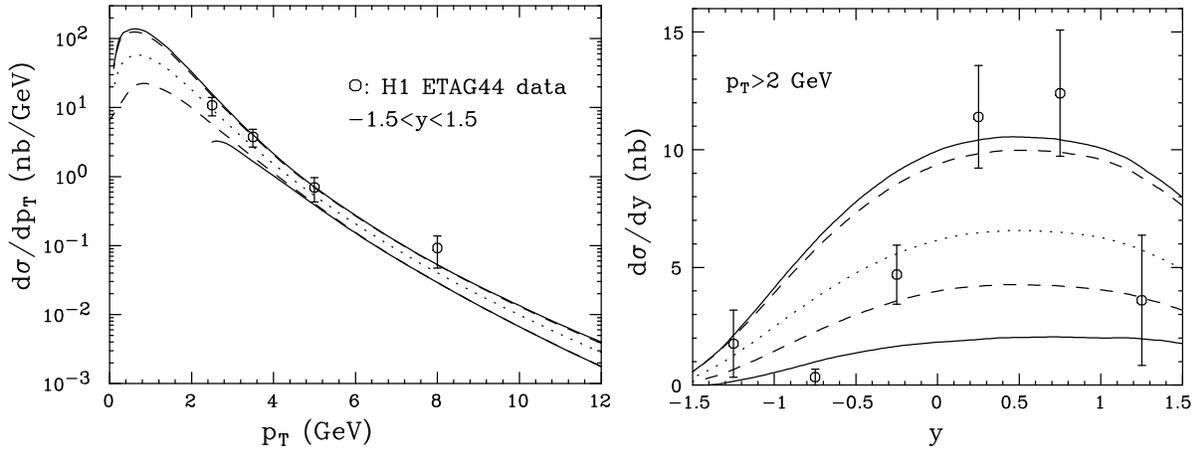
       As in fig.~\ref{fig:Zeus-Band}, for the H1 ETAG44 data.
}}
The data from H1 are compared to our calculations in fig.~\ref{fig:H33-Band},
for the ETAG33 sample, and in fig.~\ref{fig:H44-Band}, for the ETAG44 sample.
In the case of the $y$ spectra relevant to the cuts $2.5<\pt<3.5$~GeV,
$3.5<\pt<5$~GeV, and $5<\pt<10.5$~GeV, we multiply the data by the
width of the $\pt$ bin size, since they are originally quoted in
ref.~\cite{Adloff:1998vb} as $d\sigma/dyd\pt$ cross sections.
The $\pt$ spectra are perfectly described by our FONLL predictions;
all the data points are consistent with the default predictions within one 
standard deviation, with the marginal exception of the largest-$\pt$
point in the ETAG44 sample. Shape-wise, the $y$ spectrum of the ETAG33 
sample is also statistically compatible with QCD predictions; the same 
can be said for the ETAG44 sample, although in this case the statistical
significance is clearly lower.

In order to assess more quantitatively the size of the agreement
(or disagreement) between theory and data, we now present a comparison
between the two in a different form. We compute, for each data
point, the corresponding FONLL default prediction, and we consider the
ratio data/FONLL. The results are presented as open points in 
fig.~\ref{fig:rat-pt} for the $\pt$ spectra, and in 
figs.~\ref{fig:rat-eta} and~\ref{fig:rat-h44y} for the $\eta$ 
and $y$ spectra.
The figures also display bands, whose pattern is identical to that
of the bands in figs.~\ref{fig:Zeus-Band}--\ref{fig:H44-Band}. The
current bands are in fact obtained by computing the ratios of
FONLL bands over FONLL default predictions.
\FIGURE[htb]{
    \epsfig{figure=rat_pt.eps,width=0.70\textwidth}
    \caption{\label{fig:rat-pt} \protect\small
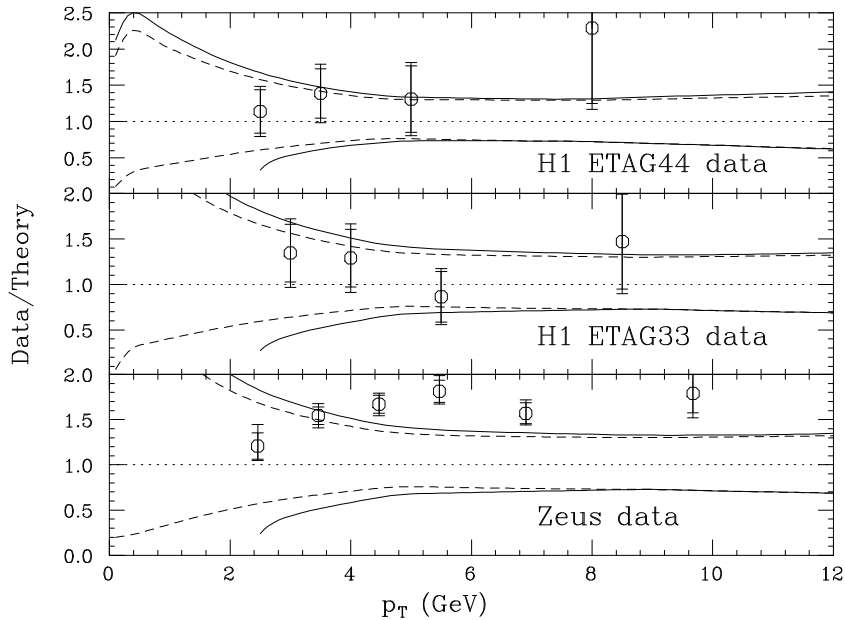
       Open points: ratio of data over FONLL default predictions,
       for transverse momentum spectra measured by Zeus and H1.
       Solid and dashed curves are the ratios of FONLL bands 
       over FONLL default predictions.
}}
\FIGURE[htb]{
    \epsfig{figure=rat_zeus_eta.eps,width=0.49\textwidth}
    \epsfig{figure=rat_h33_y.eps,width=0.49\textwidth}
    \caption{\label{fig:rat-eta} \protect\small
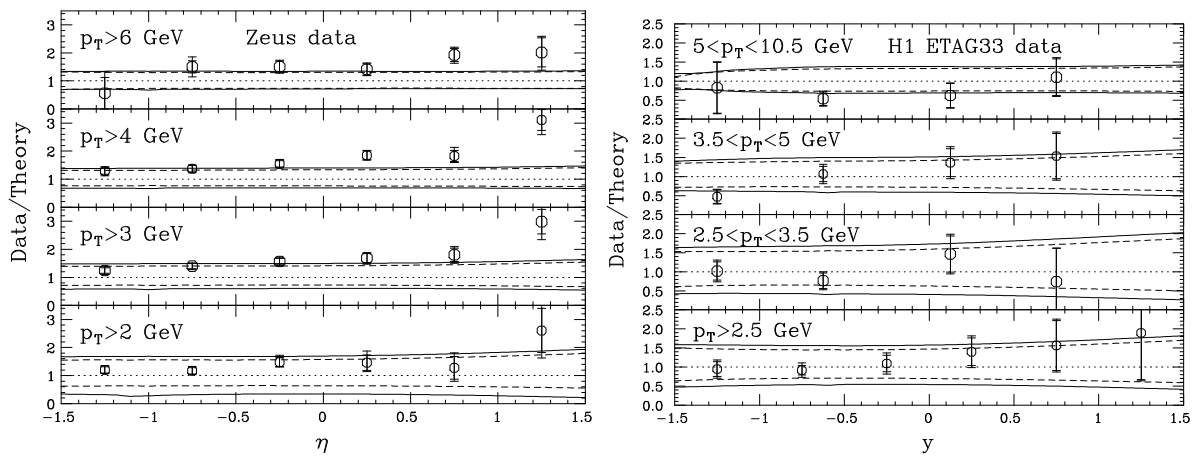
       As in fig.~\ref{fig:rat-pt}, for Zeus $\eta$ distribution
       (left) and H1 ETAG33 $y$ distribution (right).
}}
\FIGURE[htb]{
    \epsfig{figure=rat_h44_y.eps,width=0.65\textwidth}
    \caption{\label{fig:rat-h44y} \protect\small
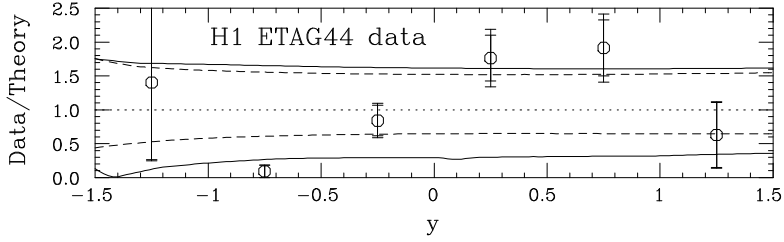
       As in fig.~\ref{fig:rat-pt}, for H1 ETAG44 $y$ distribution.
}}

The lower inset of fig.~\ref{fig:rat-pt} clearly shows that four (out
of six) data points relevant to the $\pt$ spectrum measured by Zeus are
more than one standard deviation away from the upper end of the 
theoretical band. The
measured spectrum appears to be harder than predicted; however, the
effect is probably statistically not significant in the case in which
the lowest-$\pt$ measurement is excluded from the data sample. All but
the lowest-$\pt$ point are more than 50\% away from the default prediction,
but still within a factor of 2. On the other hand, there appears to be
a nice agreement between H1 ETAG33 and FONLL predictions; ETAG44 data
are also reasonably reproduced, but there the large errors prevent from
reaching firm conclusions.

The case of the $\eta$ and $y$ spectra is dealt with in 
figs.~\ref{fig:rat-eta} and~\ref{fig:rat-h44y}. As far as the Zeus data
are concerned, fig.~\ref{fig:rat-eta} confirms the trend already 
visible in fig.~\ref{fig:Zeus-Band}: for all the $\pt$ cuts, the measured
distributions are more enhanced in the positive-$\eta$ region than FONLL
predicts. This trend is however weaker if the rightmost data points,
those affected by the largest errors, are excluded from the data sample.
Also, by excluding these points all data are within a factor of 2 from 
the default theory result. In general, most of the experimental results
lie inside the theoretical uncertainty band. As already in the case of 
the $\pt$ spectrum, H1 data do not suggest any statistically-significant
deviation from FONLL predictions.

From this study, we see that Zeus and H1 data compare differently to
FONLL predictions, the former displaying some disagreements with
theory that are not present in the latter.  It is difficult to
understand whether the two data sets are compatible within errors,
since the two experiments have different visible regions, and use
different observables (Zeus use $\eta$, H1 use $y$).

\FIGURE[htb]{
    \epsfig{figure=Zeus-pt-fo.eps,width=0.49\textwidth}
    \epsfig{figure=Zeus-eta-fo.eps,width=0.49\textwidth}
    \caption{\label{fig:Zeus-fo} \protect\small
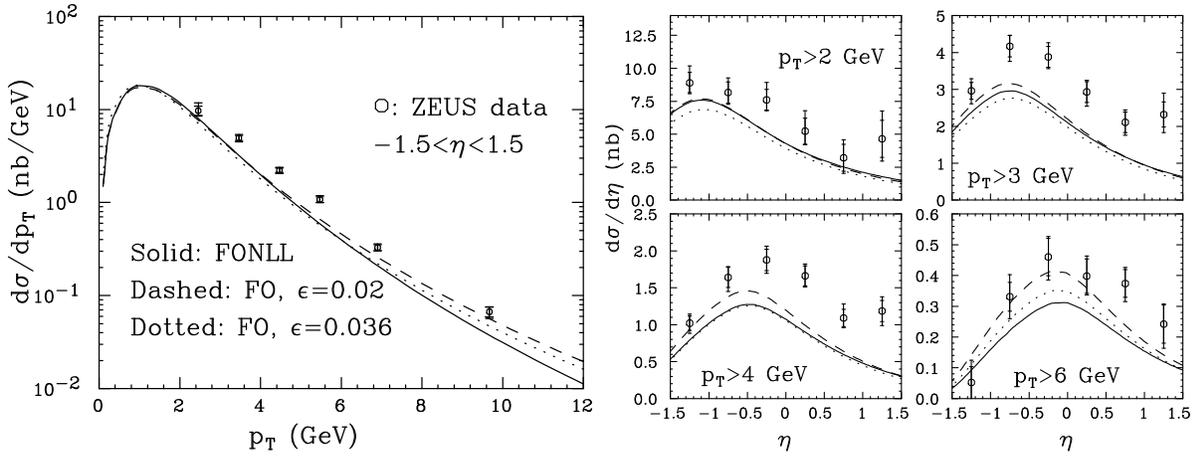
       Comparison of Zeus data with the theoretical predictions.
       The solid curve is FONLL, the dashed curve is FO (at NLO).
}}
The error bands presented in this section do not include any of
the uncertainties relevant to the non-perturbative fragmentation
discussed in sect.~\ref{sec:npff}. This appears to be justified
for the choice of the fragmentation frame and of the on-mass-shell
condition, given the size of the effects induced. The choice 
of the $\epsilon$ parameter is phenomenologically more relevant,
since it can give an up to 20\% effect on the $\pt$ spectrum.
However, it has to be kept in mind that $\epsilon$ is not
a physical quantity; in particular, its value crucially depends
upon the scheme in which the hard cross section, which is
convoluted with the NPFF, is defined. According to ref.~\cite{Nason:1999zj},
$\epsilon=0.02$ is the value to be used in the case of a FONLL
computation in order to obtain a meaningful comparison with data.
The study of sect.~\ref{sec:npff} simply suggests that a larger
(smaller) $\epsilon$ value gives a softer (harder) $\pt$ spectrum.
Thus, the comparison of FONLL predictions to Zeus data appears
to imply the necessity of an even smaller $\epsilon$ value than
the LEP data fit performed in ref.~\cite{Nason:1999zj} suggests.
On the other hand, H1 data appear to be consistent with $\epsilon=0.02$.

It is interesting to compare the FONLL and FO calculations. The two 
predictions are displayed in fig.~\ref{fig:Zeus-fo} for the Zeus cuts.
FONLL results are obtained with the default
parameters. We show two sets of FO curves. One set (dotted curves) is obtained
with default parameters except for the value of $\epsilon$, set 
equal to 0.036. This is in fact the value that gives the best
fit to $e^+e^-$ $D^*$ production data when a FO calculation
is used \cite{Nason:1999zj}.
The other set (dashed curves) has $\epsilon=0.02$. We include it in
order to emphasise, in comparison to the FONLL result, the effect 
of the large-log resummation, which is only present in the latter.
As expected (see sect.~\ref{sec:intro} and ref.~\cite{Cacciari:2001td}),
FONLL and FO predictions (with the same $\epsilon$) 
coincide at small $\pt$'s; the difference between
the two becomes visible for $\pt>4$~GeV. Where the two predictions
differ, FONLL must be considered superior. Notice that although none
of the two predictions agree well with data, the shape is better
described by the FONLL calculation. We interpret this result as
evidence of the onset of resummation effects in the data.
In fact, the FO result requires larger value of $\epsilon$ in order to
compensate for the lack of resummation effects.

\FIGURE[htb]{
    \epsfig{figure=Zeus-pt-c0.eps,width=0.49\textwidth}
    \caption{\label{fig:Zeus-c0} \protect\small
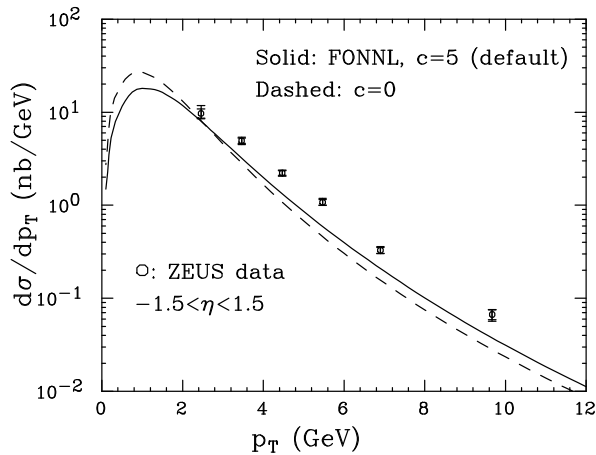
       Comparison of the $c=5$ and $c=0$ choice.}}
We do not present here RS predictions, since we believe that in
most of the visible range the transverse momentum is simply too small
for them to apply.
The RS result is below our FONLL result by 20\% to 30\% in the region
$4<\pt<20$~GeV, the difference becoming smaller as $\pt$ increases.
This difference is due to our choice of the $c$ parameter that appears
in eq.~(\ref{gfundef}), which we fixed to 5 on the basis of the fact that, in
the hadronic component of the cross section at the NLO level, the massless
limit of the cross section is a good approximation of the massive cross
section only for $\pt \gtrsim 5m$ \cite{Cacciari:2001td,Cacciari:1998it}.
We show our prediction for the extreme choice $c=0$ in fig.~\ref{fig:Zeus-c0}.
The $c=0$ choice has a softer spectrum, and it undershoots
our default choice by at most 25\% in the large $\pt$ region.

The definition of a photoproduction event in Zeus and H1
is slightly different, since the allowed photon virtuality
is much smaller in H1. Thus, one is tempted to ask whether
the differences of H1 and Zeus data compared to QCD calculations
are due to the fact that Zeus is not completely in the
photoproduction regime. The Weizs\"acker-Williams approximation is valid 
up to non-factorizable terms of order \cite{Frixione:1993yw}
\begin{equation}
\delta=\frac{Q^2_{\rm max}}{S_{\rm min}}=\frac{Q^2_{\rm max}}{4m^2}\;,
\end{equation}
which yields $\delta\approx 0.1$ for Zeus, and $\delta\approx 0.001$ for H1.
This does not mean that one can expect non-factorizable terms of order 10\%
in the Zeus case, since the leading Weizs\"acker-Williams term is
logarithmically enhanced by a factor of $\log 4m^2/m_e^2 \approx 17$.
Thus, one expects non-factorizable terms to be of the order of
a percent in the Zeus case, and one hundred times smaller in the H1 case.
It is thus unlikely that they could be the cause of the differences.
A direct comparison of Zeus and H1 data, using the same kinematical
cuts, would thus be useful to understand whether the two data sets
are consistent among each other.
\section{Conclusions}\label{sec:conc}
In this paper, we have presented comparisons between theoretical
FONLL predictions and data relevant to the photoproduction of
$D^*$ mesons, as measured at HERA by the Zeus and H1 collaborations.
We have studied the uncertainties affecting QCD predictions, and thus
estimated the largest possible range for theoretical cross sections.
Generally speaking, the agreement between theory and data is acceptable,
after a tuning of the input parameters. While H1 data do not display
any significant discrepancies with FONLL predictions, Zeus data seem
to suggest a harder transverse momentum spectrum, and an excess
towards the positive-pseudorapidity region. A detailed comparison
between the two sets of data is not possible due to the different
observables and visible regions used by the two experiments.

Having established that QCD does a reasonable job in describing
charm data, it is interesting to notice that Zeus and H1 collaborations have
now presented several results on bottom production
\cite{Adloff:1999nr,Breitweg:2000nz,blife,bDIS}, that compare much worse 
to QCD predictions than analogous charm results.
Since QCD expectations for charm agree reasonably with data, one would expect
even better agreement for the bottom quark, because of the larger
mass. It is thus unlikely that the bottom excess may be attributed to
a failure of QCD predictions.

\acknowledgments
We wish to thank the CERN theory division
for kind hospitality during the preparation of this work.
We also would like to thank M. Cacciari, L. Gladilin and
C. Grab for useful discussions.
\providecommand{\href}[2]{#2}\begingroup\raggedright\endgroup

\end{document}